\documentstyle[prb,aps]{revtex}
\begin{document}
\draft
\tighten
\twocolumn[\hsize\textwidth\columnwidth\hsize\csname @twocolumnfalse\endcsname
\title{Quantum Interference on the Kagom\'e Lattice}
\author{Yeong-Lieh Lin and Franco Nori}
\address
{Department of Physics, The University of Michigan, Ann Arbor,
Michigan 48109-1120}
\date{\today}
\maketitle
\begin{abstract}
We study quantum interference effects due to electron motion on the
Kagom\'e lattice in a perpendicular magnetic field. These effects arise
from the interference between phase factors associated with different
electron closed-paths. From these we compute, analytically and
numerically, the superconducting-normal phase boundary for Kagom\'e
superconducting wire networks and Josephson junction arrays. We use an
analytical approach to analyze the relationship between the
interference and the complex structure present in the phase boundary,
including the origin of the overall and fine structure.
Our results are obtained by exactly summing over one thousand billion billions
($\sim 10^{21}$) closed paths, each one weighted
by its corresponding phase factor
representing the net flux enclosed by each path.
We expect our computed mean-field phase diagrams to compare well with
several proposed experiments.
\end{abstract}
\vskip2pc]
\narrowtext

\section{Introduction}
Interest in  Kagom\'e structures has been fueled by several recent
developments. The first one refers to a central question in magnetism:
the $T\!=\!0$ order of the two-dimensional (2D) nearest-neighbor-coupled
Heisenberg
antiferromagnet. The Kagom\'e lattice seems to be the first 2D spin-1/2
model with vanishing further-neighbor interactions which appears to have
a disordered ground state\cite{haf}. Second, measurements of the
heat capacity of $ ^3$He absorbed on graphite at millikelvin
temperatures have been recently interpreted using a Kagom\'e lattice
structure\cite{he3}. Third, measurements on the layered oxide
SrCr$_{8-x}$Ga$_{4+x}$O$_{19}$, with Kagom\'e-like layers, are
currently attracting considerable attention\cite{scgo}.

The connection between the {\em ideal\/} Kagom\'e network and the
{\em real\/}
structures mentioned above (e.g., $ ^3$He absorbed on graphite
and SrCr$_{8-x}$Ga$_{4+x}$O$_{19}$) is somewhat unclear because
of the very important effects of
disorder, impurities, three-dimensionality, etc., present in those
materials. On the other hand, superconducting networks,
made with electron-beam lithography, offer the possibility of
experimentally studying, for the first time, nearly perfect Kagom\'e
structures. When immersed in an externally applied magnetic field,
superconducting
networks\cite{1} made of thin wires, proximity-effect junctions, and
tunnel junctions exhibit complex and interesting forms of phase
diagrams (i.e., the resistive transition temperature as a
function of the magnetic field).

The goals of this paper are to analytically study quantum interference
effects of very many ($\sim 10^{21}$) electron closed
paths on a Kagom\'e lattice
in a transverse magnetic field $B$; and from these to theoretically
predict a measurable quantity, the field-dependent
superconducting-normal phase boundary $T_{c}(B)$, for superconducting
Kagom\'e networks. The study of quantum interference effects is also
important for other physical processes. For instance, von Delft and
Henley\cite{henley} have recently investigated destructive quantum
interference between different paths that connect the same initial and
final configurations in order to study spin tunneling events on a
Kagom\'e lattice.

\section{Summary of results}
The principal results presented in this
paper are the following: (1) an analytic study of electron quantum
interference effects from sums over magnetic phase factors on closed
paths; (2) a very efficient computation of these ``discrete path
integrals" in closed-form expressions; and (3) from the latter, the
derivation of $T_{c}(B)$. The lattice path
integrals obtained here are ``many-loop" generalizations of the
standard ``one-loop" Aharonov-Bohm-type argument (where the electron
wave function picks up a phase factor $e^{i\Phi}$ each time it goes
around a closed loop enclosing a net flux $\Phi$).

The calculation of lattice path integrals will enable us to obtain the
phase boundaries\cite{niu} through an iterative approach. The spirit of
the approach follows Feynman's programme: to derive physical quantities
in terms of ``sums over paths". This method is considerably different
from the standard ones for computing $T_{c}(B)$ (e.g., numerically
diagonalizing the Ginzburg-Landau equations for large and/or irregular
structures). We will examine these issues in more detail below.

\section{Quantum interference from sums over paths}
\subsection{Physical interpretation}
The physics of
$T_{c}(B)$ is determined by the electronic {\em kinetic\/}
energy because the applied field induces a diamagnetic
current in the superconductor\cite{1}. This current (i.e., a velocity)
determines the kinetic energy of the system.
In other words, the kinetic energy can be written in terms of
the temperature as
$$-\frac{{\hbar}^{2}}{2 m^{\ast}}
\mbox{\boldmath $\bigtriangledown$}^{2} \ \sim \
-\ \frac{{\hbar}^{2}}{2m^{\ast} {\xi(T)}^2} \ \sim \ T_c(B)-T_c(0),$$
where, for any superconductor, $m^{\ast}$ is twice the electron mass,
and $\xi(T) = \xi(0) /(1-T_c(B)/T_c(0))^{1/2}$,
is the temperature-dependent coherence length.
We thus consider the electronic kinetic energy,
\begin{equation}
H=\sum_{\langle ij \rangle}c_{i}^{\dag}c_{j}\exp(i A_{ij}),
\end{equation}
on a discrete Kagom\'e lattice in a magnetic field, where
$A_{ij}=\int_{i}^{j}{\bf A}{\cdot}d{\bf l}$ is the line integral of the
vector potential along the bond from $i$ to $j$. Throughout this paper,
we set equal to one the constant factor in $A_{ij}$, namely $2\pi$
divided by the flux quantum $hc/2e$.

The lattice path integral (or moment) of order $l$, summarizing the
contribution to the electron kinetic energy of {\em all} paths
of $l$-steps, is
defined as
\begin{equation}
m_l\equiv \langle\psi_1|H^l|\psi_1\rangle,
\end{equation}
 where $|\psi_1\rangle$ denotes a
localized one-site electron state. Let us examine more closely the
physical meaning of $m_l$. The Hamiltonian $H$ is applied $l$ times to
the initial state $|\psi_1\rangle$, resulting in the new state
$H^l|\psi_1\rangle$ located at the end of the path traversing $l$
lattice bonds. Because of the presence of a magnetic field, a factor
$e^{i A_{ij}}$ is acquired by an electron when hopping through two
adjacent sites $i$ and $j$.

The above expectation value $m_l$ is non-zero
only when the path ends at the starting site. The geometric
significance of $m_l$ thus becomes clear: it is {\em the sum of the
contributions from all closed paths\/} of $l$ steps starting and ending
at the same site, each with a phase factor of $e^{i\Phi_P}$ where
$\Phi_P$ is the {\em net\/} flux enclosed by the closed path $P$,
namely
\begin{equation}
m_l= \sum_{All\ closed\ paths} \ e^{i\Phi_P}.
\end{equation}
Here, quantum interference arises because the phase factors of
different closed paths, including those from all kinds of distinct
loops and separate contributions from the same loop, interfere with
each other. Sometimes, the phases corresponding to subloops of a main
path cancel.

Several examples of different paths on a Kagom\'e lattice
and their respective contributions to the lattice path integrals are
illustrated in Figure~1. Figure~2 shows three different paths traveling
through the same loop and their separate phase factor contributions to
$m_6$. It will be seen below that the phase factors describe, in a
gauge invariant way, the electron interference effects, due to the
presence of a magnetic field, on the transition temperature; and they
are the source of the rich structure present in the phase diagram.

\subsection{Analytical calculations}
We have worked out a considerable number of lattice path integrals (up
to the $38$th order) for the Kagom\'e lattice in a uniform magnetic
field.  The first five moments can be easily computed by hand, while
the higher-order moments can be most conveniently obtained by using
recursion relations and symbolic manipulation programs. We have
manually computed moments for
the first $12$ orders. The correctness of the calculated moments is
assured by the consistency of the results obtained by hand and by
computer. The details of the computational techniques used to compute
high-order moments will be presented elsewhere.

It is instructive to explain how the first seven lattice path integrals
are obtained.  This will also clarify their physical meaning.
Let $\phi$ be the flux through an {\em elementary}
triangular cell in the Kagom\'e lattice. Since there is no path of one
step for returning an electron to its initial site, $m_1$ is always
equal to zero. There are four closed paths of two steps each
[retracing each other on one bond
($\cdot\mbox{\boldmath $\leftrightarrow$}$),
where the $\cdot$ indicates the initial site], thus
$$m_2\ =\ 4\,\cdot\!\mbox{\boldmath $\leftrightarrow$}\
=4\ e^{i0\phi}\ =\ 4=\ z,$$ where $z$ is the
coordination number of the lattice.

There are
four $3$-step closed paths enclosing a triangular cell
[two counterclockwise
($\cdot\!\!\stackrel{\leftarrow}{\mbox{\boldmath $\bigtriangledown$}}$),
like the one shown in Fig.~1(a), and two clockwise
($\cdot\!\!\stackrel{\rightarrow}{\mbox{\boldmath $\bigtriangledown$}}$)].
Thus
$$m_{3}\ =\ 2\,
\cdot\!\!\stackrel{\leftarrow}{\mbox{\boldmath $\bigtriangledown$}}\,+\,
2\,\cdot\!\!\stackrel{\rightarrow}{\mbox{\boldmath $\bigtriangledown$}}\ =\
2\,e^{i\phi}\,+\,2\,e^{-i\phi}\ =\ 4\,\cos\phi.$$

There are $28$ closed paths of
four steps each: four retracing twice on one bond
($\cdot\mbox{\boldmath $\stackrel{{\textstyle
\leftrightarrow}}{\leftrightarrow}$}$);
twelve starting from a site connecting two adjacent bonds
and retracing once on each bond
($\mbox{\boldmath $\leftrightarrow$}\!\cdot\!\mbox{\boldmath
$\leftrightarrow$}$); and twelve moving two bonds away and then
two bonds back to the original site ($\cdot\mbox{\boldmath
$\stackrel{{\textstyle \leftarrow}}
{\rightarrow}$}
\mbox{\boldmath $\stackrel{{\textstyle \leftarrow}}
{\rightarrow}$}$). Since all of them enclose no area
(i.e., no flux), then
$$m_{4}\ =\ 4\,\cdot\!\mbox{\boldmath $\stackrel{{\textstyle
\leftrightarrow}}{\leftrightarrow}$}\,+\,
12\,\mbox{\boldmath $\leftrightarrow$}\!\cdot\!
\mbox{\boldmath $\leftrightarrow$}\,+\,
12\,\cdot\!\mbox{\boldmath $\stackrel{{\textstyle \leftarrow}}
{\rightarrow}$}
\mbox{\boldmath $\stackrel{{\textstyle \leftarrow}}
{\rightarrow}$}\ =\ 28.$$

There are $60$ closed paths (30 counterclockwise and
30 clockwise) with five steps each. Three of these steps enclose a
triangular cell while the other two steps retrace each other
(e.g., $\cdot\!\!\stackrel{\leftarrow}{\mbox{\boldmath
$\bigtriangledown$}}\!\!\mbox{\boldmath $\leftrightarrow$}$,
$\cdot\!\!\stackrel{\rightarrow}{\mbox{\boldmath $\bigtriangledown$}}
\!\!\mbox{\boldmath $\leftrightarrow$}$,
$\cdot\mbox{\boldmath $\leftrightarrow$}\!\!\stackrel{\leftarrow}
{\mbox{\boldmath $\bigtriangledown$}}$,
$\cdot\mbox{\boldmath $\leftrightarrow$}\!\!\stackrel{\rightarrow}
{\mbox{\boldmath $\bigtriangledown$}}$,
$\stackrel{\leftarrow}{\mbox{\boldmath $\bigtriangledown$}}
\!\!\cdot\mbox{\boldmath $\leftrightarrow$}$, and
$\stackrel{\rightarrow}{\mbox{\boldmath $\bigtriangledown$}}
\!\!\cdot\mbox{\boldmath $\leftrightarrow$}$). Thus
$$m_{5}\ =\ 30\,e^{i\phi}\,+\,30\,e^{-i\phi}\ =\ 60\,\cos\phi.$$

Among the $6$-step closed
paths, four of them enclose an elementary hexagonal cell (like the one
shown in Fig.~1(b)), four paths go twice around the same triangle, $12$
``hourglass" paths
(\ \mbox{\boldmath $\bowtie$}\ )
surround two triangular cells ($6$ enclosing $2\phi$
and $6$ enclosing $-2\phi$), and $244$ paths enclose no {\em net} flux.
It follows then that
$$m_6\ =\ 244\,+\,16\,\cos2\phi\,+\,4\,\cos6\phi.$$

Among the
$7$-step closed paths, $28$ of them enclosing adjacent triangular and
hexagonal cells ($14$ counterclockwise and $14$ clockwise) contribute
$14e^{7 i\phi}+14e^{-7 i\phi}=28\cos7\phi$ to $m_7.$   Since a triangular
elementary cell can be traversed in $756$ different ways, with four steps
out of seven enclosing no area, it follows that
$$m_{7}\,=\,756\,\cos\phi\,+\,28\,\cos7\phi.$$

Below we present the results for $m_{8}$ through $m_{15}$, while
$m_{16}$ to $m_{20}$ are listed in Appendix A and
$m_{21}$ to $m_{38}$ will not be presented. They are
\begin{eqnarray*}
m_{8}&=&2412+416\cos2\phi+96\cos6\phi+80\cos8\phi\,,  \\
m_{9}& = &9216\cos\phi+76\cos3\phi+36\cos5\phi   \\
     &+&756\cos7\phi+120\cos9\phi\,,    \\
m_{10}& = &25804+7560\cos2\phi+1860\cos6\phi  \\
      &+&2480\cos8\phi+100\cos10\phi+20\cos14\phi\,,  \\
m_{11}& = &112420\cos\phi+2816\cos3\phi+1276\cos5\phi \\
      &+&14608\cos7\phi+4400\cos9\phi+44\cos11\phi   \\
      &+&44\cos13\phi+176\cos15\phi\,, \\
m_{12}& = &290956+1119680\cos2\phi+656\cos4\phi \\
      &+&33120\cos6\phi+51984\cos8\phi+4560\cos10\phi  \\
      &+&36\cos12\phi+1104\cos14\phi+672\cos16\phi \\
      &+&16\cos22\phi\,, \\
m_{13}& = &1385436\cos\phi+66872\cos3\phi+30680\cos5\phi  \\
      &+&248872\cos7\phi+104364\cos9\phi+2756\cos11\phi   \\
      &+&2184\cos13\phi+8372\cos15\phi+1456\cos17\phi  \\
      &+&52\cos21\phi+156\cos23\phi\,, \\
m_{14}& = &3405448+1774248\cos2\phi+31284\cos4\phi  \\
   &+&559160\cos6\phi+929320\cos8\phi+127876\cos10\phi  \\
    &+&2632\cos12\phi+35532\cos14\phi+32256\cos16\phi  \\
   &+&1960\cos18\phi+56\cos20\phi+1260\cos22\phi  \\
  &+&672\cos24\phi+28\cos30\phi\,,   \\
m_{15}& = &17292440\cos\phi+1307960\cos3\phi+627624\cos5\phi  \\
      &+&3985380\cos7\phi+2047920\cos9\phi+97860\cos11\phi   \\
      &+&68340\cos13\phi+246500\cos15\phi+74760\cos17\phi  \\
      &+&1700\cos19\phi+3420\cos21\phi+10140\cos23\phi  \\
     &+&1680\cos25\phi+120\cos29\phi+300\cos31\phi\,.
\end{eqnarray*}
Note that the even (odd) moments depend only on the even (odd)
harmonics of the flux. These moments are obtained by
summing an enormous number ($10^{21}$) of closed paths, each one
weighted by its corresponding phase factor representing the net
flux enclosed by the path.

These moments, and the $m_{21}$
through $m_{38}$ not shown, will be used in section IV to obtain
superconducting-normal phase boundary for the Kagom\'e network.

\subsection{Differences between our approach and the traditional
moments and Lanczos methods}

In electronic structure calculations there is a method to compute the
density of states called the moments method.  This is similar to our
approach in the sense that both compute moments.  However, also in
statistical analysis moments are often used. Moments are simply a concept of
wide applicability to a large number of dissimilar problems.   For
instance, the half-width half-maximum of a probability distribution
can be expressed in terms of its second moment; the center of mass of
a distribution can be expressed in terms of its first moment; and
the skewness or degree of asymmetry in terms of its
third moment.  Any mathematical function (e.g., probability distribution,
Raman lineshape, or any other function) can always be expressed in
terms of an expansion using an infinite number of moments; although
typically the first ten or so moments provide an excellent
approximation to most functions.  In our system, we have analytically
computed moments to extremely high order (the first $38$ moments).

We list several important differences between the
standard ``moments method''
and our problem.  The typical use of the moments method:
$(i)$ focuses on the computation of electronic density of
states (instead of superconducting $T_c$'s);
$(ii)$ is totally {\it numerical\/} (instead of mostly analytical);
$(iii)$ is done at zero magnetic field (instead of obtaining
expressions with an explicit field dependence);
$(iv)$ does not focus on the explicit computation of path integrals; and
$(v)$ does not study the physical effects of quantum interference
(which is at the heart of our calculation and physical interpretation).

In conclusion, the traditional use of the moments method in condensed matter
is significantly different from the approach and problem studied here.

Another way to diagonalize Hamiltonians is called the Lanczos method. This
method directly obtains the tridiagonal form, {\it without\/}
computing the moments; thus differs in a significant way from the approach
used here (where the explicit computation of the moments is one of our
goals, since they can be used for other electronic property calculations).
\ Furthermore, it is not convenient to use standard Lanczos method in our
particular problem because it is extremely difficult to directly derive the
parameters and the states of the iterative tridiagonalization
procedure.  This is so because of the presence of the magnetic
field.  On the other hand, the moments method provides
standard procedures to diagonalize a matrix {\it after\/} the moments
are computed.

Our approach has features in common with the Lanczos method and features
in common with the moments method (e.g., the
computation of the moments, which is absent in the canonical Lanczos
technique).  More importantly, the five differences listed above
($(i)-(v)$) make our approach to this problem quite different to
the standard (purely numerical) implementations of both the Lanczos
and moments methods.

\section{Kagom\'e superconducting  networks and Josephson junction
arrays}

The complex structure in $T_c(B)$ is essentially a result of
quantum interference effects due to the electron kinetic energy and the
multi-connectedness of the networks in a magnetic field.  The magnetic
fluxes through the elementary cells are useful parameters to
characterize the interference effect. At zero magnetic field, the
quantum interference effect is absent, and therefore the resistive
transition temperature should have a peak.

Mean field theory\cite{1,niu,shih} is very effective in providing a
quantitative description of the phase diagrams. For wire networks, the
mean field expression is given by the Ginzburg-Landau equation
expressed in terms of the order parameters at the nodes. For a junction
array, one has a set of self-consistent equations\cite{shih} for the
thermally averaged pair wavefunctions of the grains.  Such equations
are linearized near the transition point, and the
highest temperature (i.e., top eigenvalue)
at which a non-trivial solution first appears is identified as the
transition temperature. This can be seen because
the kinetic energy of the system can be written in terms of
the temperature as $-{\hbar}^{2}/(2m^{\ast} {\xi(T)}^2)
\sim T_c(B)-T_c(0).$ Therefore, one is left to find the top
spectral edge of an eigenvalue problem\cite{1,niu,shih}.  In summary,
these equations can be mapped into a tight-binding Schr\"{o}dinger
problem for an electron hopping on
a lattice immersed in a magnetic field and $T_c(B)$ is determined by
the kinetic energy of the electrons (i.e., by the Hamiltonian $H$).

Now we are going to find the top spectral edge, which is proportional
to the transition temperature.
The best strategy for deriving eigenvalues and eigenvectors
of symmetric matrices is to first reduce the matrix to some simple form,
which is typically tridiagonal.  All methods employed today
(almost exclusively numerical, and not analytical as done here)
are based on modifications of the Jacobi method.  The latter has been
in widespread use long time before the Lanczos and moments methods
were developed.

We choose a normalized initial electron
state $|\psi_1\rangle$, which is strongly localized at one site,
and perform the following expansion:
\begin{equation}
H|\psi_n\rangle=\gamma_{n}|\psi_{n-1}\rangle+\beta_{n}|\psi_n\rangle
+\gamma_{n+1}|\psi_{n+1}\rangle,
\end{equation}
with the condition $|\psi_{0}\rangle\equiv0$. The $H$-matrix in the
basis $|\psi_n\rangle$ is obviously in a real tridiagonal form. The
expansion is useful because finite truncations give good approximations
to the quantity we desire, i.e., the top spectral edge.

Each new state
in this method expands outward by one more step from the site where the
starting state is located.  Thus an $n$th-order truncation can cover a
region of radius $n$ on the network\cite{niu}.  Furthermore,
the parameters in
the truncated Hamiltonian are gauge-invariant quantities.

The non-zero matrix elements, $\beta$'s and $\gamma$'s, can be exactly
expressed in terms of the moments of the Hamiltonian by using the
following novel and systematic procedure. First, define the auxiliary
matrix $M$ with the first row elements given by $M_{0,l}\equiv m_{l}$.
The other rows are evaluated by using only one immediate predecessor
row through
\begin{eqnarray}
M_{n,l}&=&\frac{M_{n-1,l+2}-M_{n-1,1}\,M_{n-1,l+1}}{M_{n-1,2}
-M_{n-1,1}^{2}}  \nonumber  \\
& & \mbox{}-\sum_{k=0}^{l-1}\ M_{n,k}\,M_{n-1,l-k},
\end{eqnarray}
where $n$ and $l \geq 1$. The $\beta_{n}$'s and $\gamma_{n}$'s
are obtained from the elements
of the second and third columns as
\begin{equation}
\beta_{n}=M_{n-1,1}
\end{equation}
and
\begin{equation}
\gamma_{n}=\sqrt{M_{n-1,2}-M_{n-1,1}^{2}}.
\end{equation}
Note that elements in the first column
$M_{n,0}$ are always equal to 1.

Equations (5)-(7), which are not present in Ref.~6,
allow a more systematic computation of the tridiagonal elements $\beta_n$
and $\gamma_n$.  Furthermore, only a few moments were computed in Ref.~6,
and none of them applied to the Kagom\'e lattice.  The much higher-order
moments computed here (involving billions of billions of paths), coupled
with the systematic evaluation of the parameters $\beta_n$ and $\gamma_n$,
give an unprecedented level of accuracy in the derivation of the structure
in $T_c(\phi)$.

Below we explicitly
express the first few $\beta$ and $\gamma$ parameters in terms of
the moments. They apply to {\em any} type of lattice.
\begin{eqnarray*}
\beta_1&\,=\,&0\,, \\
\beta_2&\,=\,&\frac{m_3}{z}\,,   \\
\beta_3&\,=\,&\frac{m_{5}z^{2}-2m_{4}m_{3}z+m_{3}^{3}}{z\alpha}\,,
\end{eqnarray*}
and
\begin{eqnarray*}
\gamma_1&\,=\,&\sqrt{z}\,,  \\
\gamma_2&\,=\,&\frac{\sqrt{\alpha}}{z}\,, \\
\gamma_3&\,=\,&\frac{\sqrt{zA}}{\alpha}\,,
\end{eqnarray*}
where
$$\alpha\,=\,m_{4}z-m_{3}^{2}-z^{3},$$
and
\begin{eqnarray*}
A&\,=\,&m_{6}m_{4}z-m_{6}m_{3}^{2}-m_{6}z^{3}+2m_{5}m_{4}m_{3}
+2m_{5}m_{3}z^{2} \\
&\,-\,&m_{5}^{2}z-3m_{4}m_{3}^{2}z+m_{4}^{2}z^{2}-m_{4}^{3}
+m_{3}^{4}\,.
\end{eqnarray*}

Using the lattice path integrals obtained, the
Hamiltonian matrix elements ($\beta$ and $\gamma$) can
be readily computed. For instance, the second order
truncation of the Hamiltonian is
$$H^{(2)}\,=\,\left( \begin{array}{cc}
0 & \sqrt{m_2} \\
\sqrt{m_2} & m_3/m_2
\end{array} \right)
\,=\,\left ( \begin{array}{cc}
0 & 2 \\
2 & \cos\phi
\end{array} \right).$$
The presence of the $\cos\phi$ matrix element corresponds (physically)
to the Little-Parks oscillations observed in a {\em single}
superconducting loop.  Its corresponding top eigenvalue is
\begin{eqnarray*}
T^{(2)}_{c}(\phi)&=&\frac{1}{2}\,\left[\ \frac{m_3}{m_2}+
\sqrt{\left(\frac{m_3}{m_2}\right)^2+4m_2}\ \right] \\
&=&\frac{1}{2}\,\left(\ \cos\phi+\sqrt{\cos^{2}\phi+16}\ \right).
\end{eqnarray*}
The other eigenvalue is not physically important
because it does not cross $T^{(2)}_{c}(\phi)$. $T^{(2)}_{c}(\phi)$
contains information on the quantum interference effects produced by
eight paths (four from $m_2$ and four from $m_3$). These closed paths,
with at most three steps, cover a very small region of the lattice.
The corresponding $\Delta T^{(2)}_{c}(\phi)$ is the top curve
shown in Figure~3.

The third order truncation of the Hamiltonian is
\begin{eqnarray*}
\lefteqn{H^{(3)}} \\
& = &
\left ( \begin{array}{ccc}
0 & \sqrt{m_2} & 0  \\
\sqrt{m_2} & m_3/m_2 &
\sqrt{\frac{m_4}{m_2}-(\frac{m_3}{m_2})^2-m_{2}} \\
0 & \sqrt{\frac{m_4}{m_2}-(\frac{m_3}{m_2})^2-m_{2}} &
\frac{m_{5}m_{2}^{2}-2m_{4}m_{3}m_2+m_{3}^{3}}
{m_4m_{2}^{2}-m_{3}^{2}m_2-m^{4}_{2}}
\end{array} \right) \\
& = & \left ( \begin{array}{ccc}
0 & 2 & 0  \\
2 & \cos\phi & \sqrt{(5-\cos2\phi)/2} \\
0 & \sqrt{(5-\cos2\phi)/2} &
\frac{7\cos\phi+\cos3\phi}{10-2\cos2\phi}
\end{array} \right).
\end{eqnarray*}
Its corresponding top eigenvalue is
$$T^{(3)}_{c}(\phi)=\frac{8\cos\phi}{3\lambda}+
\frac{31-11\cos2\phi}{3\lambda\Theta^{\frac{1}{3}}}+
\frac{32(1+\cos2\phi)}{9\lambda^2\Theta^{\frac{1}{3}}}
+\Theta^{\frac{1}{3}},$$
where
$$\lambda=5-\cos2\phi,$$ and
\begin{eqnarray*}
\lefteqn{\Theta=} \\
& &\frac{512\cos^{3}\phi}{27\lambda^3}+
\frac{102\cos\phi-22\cos3\phi}{3\lambda^2}
-\frac{7\cos\phi+\cos3\phi}{\lambda} \\
& &\mbox{}+\frac{\sqrt{3}}{9\lambda^2}(54\cos^52\phi-2285\cos^42\phi
+17832\cos^32\phi \\
& &\mbox{}-72982\cos^22\phi+171866\cos2\phi-179509)^{\frac{1}{2}}.
\end{eqnarray*}
We have analytically obtained all the eigenvalues of $H^{(3)}$.
However, the top one is the only one with physical significance
to our problem, because the two other eigenvalues (not shown here)
do not cross it. $T^{(3)}_{c}(\phi)$
contains terms up to $10\phi$, and the
corresponding  $\Delta T^{(3)}_{c}(\phi)$ is
shown in Figure~3 (curve labeled by $3$).
$T^{(3)}_{c}(\phi)$ exhibits a few terms which follow the Little-Parks
$\cos\phi$-type oscillations and also many additional
contributions that go well beyond the single-loop effect since they are
rigorously derived by summing $96$ closed-paths on the lattice.

The fourth-order truncation is given by
$$H^{(4)}\!=\!\left( \begin{array}{cccc}
0 & 2 & 0 & 0 \\
2 & \cos\phi & \sqrt{(5-\cos2\phi)/2} & 0 \\
0 & \sqrt{(5-\cos2\phi)/2} &
\frac{7\cos\phi+\cos3\phi}{10-2\cos2\phi}& \gamma_4 \\
0 & 0 & \gamma_4 & U/V
\end{array} \right),$$
here
$$\gamma_4=\sqrt{\frac{132-64\cos2\phi-6\cos4\phi+20\cos6\phi-2\cos8\phi}
{51-20\cos2\phi+\cos4\phi}},$$
\begin{eqnarray*}
U& = &344\cos\phi-109\cos3\phi-240\cos5\phi \\
   &+&570\cos7\phi-124\cos9\phi+7\cos11\phi,
\end{eqnarray*}
and
\begin{eqnarray*}
V& = &692-449\cos2\phi-8\cos4\phi \\
& + &104\cos6\phi-20\cos8\phi+\cos10\phi.
\end{eqnarray*}
We have also analytically obtained {\em all} the eigenvalues of $H^{(4)}$.
They exhibit terms which follow the
$\cos\phi$-type oscillations and also very many additional
contributions which are rigorously derived by
summing $1144$ closed-paths on the lattice.

In general, $T^{(n)}_{c}(\phi)$ is obtained from the following moments:
$m_2$, $m_3$, $\ldots$, and $m_{2n-1}$. These contain information on
the quantum interference effects due to closed paths of $2n-1$ steps.

In the following, we physically analyze how the approximations work for
the phase diagrams and how the local geometries of the lattices affect
the structures in them. Figure~3 shows the second through eighth, and
also the tenth, fifteenth, and nineteenth order approximants to the
phase diagram. The fine structures that appear in the lower order
approximants become sharper when the order is increased. Eventually,
the dips become cusps. Since the $15$th to $19$th approximants are
essentially identical, close convergence to the infinite system size
has been achieved. The cusps at $\phi=1/8$, $1/4$, $3/8$, $5/8$, $3/4$
and $7/8$ are well established in the highest order shown.  The
resulting phase diagram originates from a competition between phase
coherence at different length scales, which can be explicitly read
from the corresponding expressions of the moments.

Quantitatively, there are several differences between the $T_c(\phi)$
for the Kagom\'e and, for instance, its perhaps most similar Bravais
lattice:  the triangular network.  The following simple argument
illustrates a clear quantitative difference.  For the triangular
lattice, the energy is minimized when $\phi = n \Phi_0$, where $n$ is
an integer, $\phi $ is the flux through an elementary triangular cell,
and $\Phi_0$ is equal to $2\pi$ times the quantum of flux.  For the
Kagom\'e case this is also true: when $\phi = \Phi_0$, then $\phi_h = 6
\Phi_0$, where $\phi_h $ is the flux enclosed by an elemental hexagon.
However, when $\phi = \Phi_0 / 6$, the hexagons are in their lowest
energy configuration, but the triangles are not.  Thus, this particular
value of the flux  ($\phi=\Phi_0/6$) minimizes the energy of the
hexagonal elementary plaquettes in the Kagom\'e lattice and {\it
none\/} of the elementary cells in the triangular lattice.  This
difference will be reflected in the values for $T_c$ at $\phi =
\Phi_0/6$.

Another difference between the phase boundaries of the Kagom\'e and other
regular lattices (e.g., square, triangular, hexagonal) is that the latter
{\it all\/} have a prominent minima when the flux per elementary plaquette
is equal to $\Phi_0/2$, while the Kagom\'e does not.

This concludes our calculation and the main scope of our paper.  In the next
two sections, we briefly discuss experimental issues and some relations to
the magnetic ground state of the XY Kagom\'e Hamiltonian.

\section{Heisenberg Antiferromagnetism versus Superconductivity in
Kagom\'e Lattices}

In superconducting wire networks and arrays of Josephson-coupled islands,
the measured $T_{c}(\phi)$ is often obtained from the maximum (or some other
nearby) value of $dR/dT$, where $R$ is the resistance; i.e.,
the measured $T_{c}$ is close, but not
necessarily equal, to the real $T_c$.  This measured
$T_c$ can be theoretically
computed with good accuracy by solving linearized mean-field equations
(instead of solving, for instance, the original XY Hamiltonian for the
Josephson-coupled islands). Mean-field is appropriate because
fluctuations
are not important as long as the measurements are not extremely close
to the real $T_c$; which is the typical case.
Also, by far the dominant effect on any given superconducting island is
given by the nearest neighboring islands, and mean-field considers an
average over them. A better approximation (called cluster mean-field)
consists in considering a thermal average over further neighbors\cite{niu}.
This procedure incorporates a
significant amount of thermal fluctuations.  Another approach to
incorporate fluctuations uses real-space renormalization-group
techniques\cite{niu}.
Even the simplest mean-field for $N$ coupled islands gives $N$ coupled
nonlinear complex differential equations which in general are difficult to
solve.  Linearization simplifies the mathematical treatment of the
problem and is also a very physical consideration because
the density of Cooper pairs (or order parameter) is small near the
transition.  Thus, $T_c$ can be conveniently obtained from the ground
state energy of this linearized mean field model.

After all these steps (i.e., mean-field and linearization), it is
not clear that all the information present in the original (XY Hamiltonian)
problem can be observed in a narrow part of phase space, i.e., close to
the phase boundary.  Even if the original
information is still there, it is not clear to us that $T_c$ is the best
way to access it.  For example, the steps in between, from the (initial)
full XY Hamiltonian to the (final) linearized mean-field equations, might
reduce the symmetry of the problem; and some of the essential information
will be lost; or still there but difficult to unmask in a transparent and
quantitative way.

The antiferromagnetic XY model on the Kagom\'e lattice has a degenerate
ground state, and $T_c$ does not seem to reflect this feature because
$T_c$ does not contain all the information that can be obtained from the
full XY model (or the full nonlinear Ginzburg-Landau equations), but only
part of it.  Thus, it is not obvious to us how the degeneracy of the
ground state of the antiferromagnetic XY model influences Tc.

This paper does {\it not\/} focus on degeneracy issues, which are beyond
the scope of this work, but on a many-loop version of the Aharonov-Bohm-type
argument to derive the origin of the phase boundary features exhibited by
superconducting networks.   This approach involves moving electrons around
loops threaded by a magnetic field, and examining the effect of this
electron transport on both the electronic wave functions and ground state
energy ($T_c$).

\section{Suggested Experiments on Kagom\'e Superconducting Networks}

A measurement of $\Delta T_{c}(\phi)$ on a Kagom\'e lattice would
provide a phase boundary with features which are different from
the ones obtained in all the other networks studied so far (e.g.,
the absence of a prominent minimum at $\Phi_0/2$ and other
features discussed in section IV). More importantly, the Kagom\'e
phase boundary has many sharp dips, unlike, e.g., the relatively
dull and featureless phase diagram for the hexagonal network.
It appears to us that the dips (cusps) in the Kagom\'e
$\Delta T_{c}(\phi)$ are the {\em sharpest} of all regular lattices.
These sharp dips are predicted\cite{niu} to become even sharper when the
measurement are made closer to the real $T_c$. Furthermore, the current
state of the technology would allow the construction of lattices which
are far more perfect than the ones previously built. Thus, a direct
comparison between measurements and our Figure~3 could be made.

The Heisenberg and the XY Kagom\'e antiferromagnets have attracted
considerable attention for the reasons mentioned in our opening
paragraph and discussed in detail in references 1-3.  These models
exhibit geometrical frustration which prevents the establishment
of long-range magnetic order.  The classical Hamiltonians exhibit
a complicated ground-state degeneracy, with both noncoplanar and
planar states in the degenerate manifold.

In order to look for possible manifestations of the degeneracy and
extensive entropy of the
antiferromagnetic XY ground state, experiments could perhaps
focus also away from $T_c$. At lower temperatures, the imaging of
the vortex lattice could provide a picture of the XY spin-state,
because the
phases of the order parameters would be arranged as the XY spin
configuration of a Kagom\'e antiferromagnet.  At lower temperatures,
the nonlinearity (e.g., quartic term in the Ginzburg- Landau free
energy) breaks the degeneracy present in the linear approximation.
When lowering the temperature, the density of Cooper pairs increases
and it is not possible to linearize the equations.  Thus, the full
Hamiltonian must be used.

The variation of the phases (of the order
parameter) between nearby islands will produce loop currents and their
associated fields could, in principle, be measured using a variety of
different probes like:  $(a)$ a scanning tunneling microscope,
$(b)$ a scanning atomic force microscope, $(c)$ a scanning Hall probe,
and/or $(d)$ inductive measurements like the ones employed by Martinoli's
group.\cite{1}
Thus, information on the currents and/or flux configurations
would be helpful
to determine the phase arrangement (XY spins on a Kagom\'e lattice).

\section{Discussions}
The iterative approach presented above can be
thought of as an analytical version of a hybrid between modified Lanczos
and moments methods, both
typically used in purely numerical procedures. It not only
provides us with a systematic approximation scheme for the phase
diagrams, but also a rather powerful tool for a qualitative and
quantitative analysis of the structures in them and their physical
origin.  As the order of approximation is increased, more
geometrical information of the lattice is included in the interference
treatment, and more fine structures are resolved.  The
correspondence is such that we can draw a number of specific
conclusions regarding how the dips of various sizes in $T_c(B)$ are
related to the geometry of the underlying lattice.  The gross structure
is determined by the interference effects arising from few loops.  The
secondary dips or peaks are determined by the additional information
provided by interference arising from higher-order lattice path
integrals (i.e., larger paths that explore more loops).  In general,
higher-order fine structures are due to interference among the cells of
larger clusters (here including as many as $10^{21}$ loops).
The indexing of the dips
($\phi=n/8,$ where $n=1, 2, 3, 5, 6,$ and $7$) in
the phase diagrams can be easily read off from the periods of the
truncated matrix elements.  The sharpness of the dips is due to {\em
long range} correlations, and cannot be accounted for by local
geometries. We expect our predictions to compare well with future
experiments on this system.

\acknowledgments
It is a pleasure to acknowledge conversations with P. Chandra.
FN was supported in part from a GE fellowship,
a Rackham grant, the NSF through grant DMR-90-01502,
and SUN Microsystems.

\newpage
\onecolumn
\widetext
\appendix
\section{Lattice path integrals: $m_{16}$ through $m_{20}$}

\begin{eqnarray*}
m_{16}& = &41001292+25428480\cos2\phi+918560\cos4\phi+
9092352\cos6\phi+15316256\cos8\phi+2863616\cos10\phi  \\
   &+&  109088\cos12\phi+890560\cos14\phi+956304\cos16\phi+
112000\cos18\phi+5120\cos20\phi+53568\cos22\phi  \\
  &+&42912\cos24\phi+2688\cos26\phi+192\cos28\phi+2752\cos30\phi
+1440\cos32\phi+64\cos38\phi\,,   \\
& & \ \\
m_{17}& = &218660460\cos\phi+ 23053020\cos3\phi+11763116\cos5\phi
      +61614120\cos7\phi+36265760\cos9\phi+2637380\cos11\phi   \\
      &+&1726928\cos13\phi+5807812\cos15\phi+2339200\cos17\phi
      +113832\cos19\phi+135388\cos21\phi+385492\cos23\phi  \\
     &+&112200\cos25\phi+2992\cos27\phi+9180\cos29\phi+23052\cos31\phi
+4080\cos33\phi+340\cos37\phi+748\cos39\phi\,, \\
& & \ \\
m_{18}& =&504967204+358082904\cos2\phi+21570528\cos4\phi+
143901532\cos6\phi+240951024\cos8\phi+56411064\cos10\phi  \\
   &+& 3381816\cos12\phi+19345860\cos14\phi+22651848\cos16\phi+
3827416\cos18\phi+256560\cos20\phi+1688400\cos22\phi  \\
  &+&1610136\cos24\phi+195084\cos26\phi+17064\cos28\phi+136752\cos30\phi
+105120\cos32\phi+7560\cos34\phi \\
&+&720\cos36\phi+7668\cos38\phi+168\cos46\phi+12\cos54\phi\,,   \\
& & \ \\
m_{19}& = &2799941004\cos\phi+382148520\cos3\phi+208697444\cos5\phi
      +932000464\cos7\phi+603749320\cos9\phi   \\
      &+&60413692\cos11\phi+38512392\cos13\phi+120342732\cos15\phi
    +57911316\cos17\phi
      +4436880\cos19\phi  \\
     &+&4161456\cos21\phi+11250660\cos23\phi+4359132\cos25\phi
+246088\cos27\phi+408804\cos29\phi \\
&+&1005176\cos31\phi+303240\cos33\phi+10336\cos35\phi+30172\cos37\phi
+67108\cos39\phi \\
&+&12540\cos41\phi+1064\cos45\phi+2128\cos47\phi+76\cos53\phi
+152\cos55\phi\,, \\
& & \ \\
m_{20}& =&6338429028+4999672400\cos2\phi+445104760\cos4\phi+
2233225680\cos6\phi+3683951360\cos8\phi \\
&+&1025139280\cos10\phi+87649740\cos12\phi+383654240\cos14\phi
+471275520\cos16\phi+102042960\cos18\phi \\
&+&9360020\cos20\phi+44470640\cos22\phi+46514880\cos24\phi+8109600\cos26\phi+883500\cos28\phi+4886400\cos30\phi \\
&+&4442160\cos32\phi+595120\cos34\phi+69960\cos36\phi+
+438800\cos38\phi+328240\cos40\phi+26400\cos42\phi \\
&+&2800\cos44\phi+24080\cos46\phi+12320\cos48\phi+ 200\cos52\phi+
2080\cos54\phi+880\cos56\phi+80\cos62\phi\,.
\end{eqnarray*}

\newpage
\narrowtext
\twocolumn

\begin{figure}
\caption{Several examples of closed paths on a Kagom\'e lattice and
their respective phase factor contributions to the lattice path
integrals. (a) Traversing counterclockwise  a $3$-step path on an
elementary triangular cell  contributes a term $e^{i\phi}$ to $m_3$;
(b) a $6$-step clockwise walk on an elementary hexagonal loop
contributes a term $e^{-6i\phi}$ to $m_6$; and (c) a closed path of
$26$ steps with several links retraced. Its phase factor is
$e^{-16i\phi}.$ The solid dots denote the starting (and ending) sites
and the arrows specify the hopping directions on bonds.}
\label{fig1}\end{figure}

\begin{figure}
\caption{An ``hourglass" loop made of two triangles can be traversed in
eight different ways, contributing three different phase factors to
$m_6$. On the right-hand side, there are three possible different
paths: $\Gamma_1$, $\Gamma_2$ and $\Gamma_3$. Path $\Gamma_1$
($\Gamma_2$) contributes $e^{i\phi+i\phi}=e^{2i\phi}$
($e^{-i\phi-i\phi}=e^{-2i\phi}$) to $m_6$. In path $\Gamma_3$ the
phases cancel and the contribution of this path to $m_6$ is
$e^{i\phi-i\phi}=1$. The coefficients in front of the $\Gamma$'s are
the number of paths with the same phase factor.}
\label{fig2}\end{figure}

\begin{figure}
\caption{Superconducting transition temperature,
$\Delta T^{(n)}_{c}(\phi)=T_{c}(0)-T_{c}(\phi)$, for
a Kagom\'e lattice, versus the
magnetic flux, $\phi$, through an elementary triangular cell, computed
from the truncated Hamiltonian. From top to bottom, the orders of
truncation are $n=2$ (top curve), $3$, $4$, $5$, $6$, $7$, $8$, $10$,
$15$, and $19$.  Note the development of fine structures and cusps. The
convergence is monotonic.}
\label{fig3}
\end{figure}


\begin{references}
%
%
\bibitem{haf}P. Chandra, P. Coleman and I. Ritchey, J. Physique
(Paris) {\bf 3}, 591 (1993);
P. Chandra and P. Coleman, Les Houches Lectures, in press (1991);
P. Chandra and P. Coleman, Phys. Rev. Lett. {\bf 66}, 100 (1991);
A. Chubukov, {\em ibid} {\bf 69}, 832 (1992);
E. F. Shender, V. B. Cherepanov, P. C. W. Holdsworth,
and A. J. Berlinsky, {\em ibid} {\bf 70}, 3812 (1993);
K. Yang, L. K. Warman, and S. M. Girvin, {\em ibid} {\bf 70},
2641 (1993); C. Zeng and V. Elser, Phys. Rev. B {\bf 42}, 8436 (1990);
P. W. Leung and V. Elser, {\em ibid} {\bf 47}, 5459 (1993);
A. B. Harris, C. Kallin and A. J. Berlinsky, {\em ibid} {\bf 45},
2899 (1992);
T. C. Hsu and A. J. Schofield, J. Phys. Condens. Matter {\bf 3},
8067 (1991).
\bibitem{he3}V. Elser, Phys. Rev. Lett. {\bf 62}, 2405 (1989); D. S.
Greywall and P. A. Busch, {\em ibid} {\bf 62}, 1868 (1989); {\em ibid}
{\bf 65}, 2788 (1990).
\bibitem{scgo}A. P. Ramirez, G. P. Espinoza and A. S. Cooper, Phys. Rev.
Lett. {\bf 64}, 2070 (1990); C. Broholm, G. Aepli, G. P. Espinosa and
A. S. Cooper, {\em ibid.} {\bf 65}, 3173 (1990);
I. Ritchey, P. Coleman and P. Chandra, Phys. Rev. B {\bf 47}, 15342 (1993).

\bibitem{1} See, for instance the reviews by, B. Pannetier, J. Chaussy
and R. Rammal, Jpn. J. Appl. Phys. {\bf 26}, Suppl. 26-3, 1994 (1987);
P. Martinoli, {\it ibid}, 1989 (1987);
C. J. Lobb, Physica {\bf 126B}, 319 (1984); and the many articles
in the issue on ``Coherence in superconducting networks", J. E. Mooij
and G. B. J. Sch\"{o}n, eds, Physica {\bf 152B}, 1 (1988).

\bibitem{henley} J. von Delft and C. L. Henley, Phys. Rev. Lett. {\bf 69},
3236 (1992).

\bibitem{niu}  Q. Niu and F. Nori, Phys. Rev. B {\bf 39}, 2134 (1989).

\bibitem{shih} W. Y. Shih and D. Stroud, Phys. Rev. B {\bf32}, 158 (1985).
\end{references}
\end{document}